\documentclass{amia}
\usepackage{graphicx}
\usepackage[labelfont=bf]{caption}
\usepackage[superscript,nomove]{cite}
\usepackage{color}
\usepackage{todonotes}
\usepackage{wrapfig}
\usepackage{hyperref}
\usepackage{caption}
\usepackage{subcaption}
\usepackage{amsmath}
\usepackage{amssymb}

\begin{document}

\title{Comparing Suicide Risk Insights derived from Clinical and Social Media data}

\author{Rohith K. Thiruvalluru*, MS$^{1}$, Manas Gaur*, MS$^{2}$, Krishnaprasad Thirunarayan, PhD$^{3}$, Amit Sheth, PhD$^{2}$, Jyotishman Pathak, PhD$^{1}$}

\institutes{
    $^1$Department of Population Health Sciences, Weill Cornell Medicine, USA; \\
    $^2$Artificial Intelligence Institute, University of South Carolina, USA; \\
    $^3$Kno.e.sis Center, Wright State University, USA
    \\
    }

\maketitle

\noindent{\bf Abstract}

\textit{Suicide is the $10^{th}$ leading cause of death in the US and the $2^{nd}$ leading cause of death among teenagers. Clinical and psychosocial factors contribute to suicide risk (SRFs), although documentation and self-expression of such factors in EHRs and social networks vary. This study investigates the degree of variance across EHRs and social networks. We performed subjective analysis of SRFs, such as self-harm, bullying, impulsivity, family violence/discord, using $>$\mbox{13.8 Million} clinical notes on 123,703 patients with mental health conditions. We clustered clinical notes using semantic embeddings under a set of SRFs. Likewise, we clustered 2180 suicidal users on r/SuicideWatch ($\sim$30,000 posts) and performed comparative analysis. Top-3 SRFs documented in EHRs were depressive feelings (24.3\%), psychological disorders (21.1\%), drug abuse (18.2\%). In r/SuicideWatch, gun-ownership (17.3\%), self-harm (14.6\%), bullying (13.2\%) were Top-3 SRFs. Mentions of Family violence, racial discrimination, and other important SRFs contributing to suicide risk were missing from both platforms.}

\section*{Introduction} \label{sec:I}
Suicide is one of the leading causes of death in the US \cite{Tang2020}. With an estimated increase of 61\% in mental health patients per mental healthcare providers (MHPs) by 2025, it is hard to maintain patient engagement with treatment \cite{Callahan2013,Y.2017}.  Further, predicting when someone will attempt suicide has been nearly impossible. 
Prior research has identified suicidal behavior using health insurance claims and electronic health record (EHR) data \cite{Callahan2013, Y.2017}. It is widely recognized that such data may have low sensitivity to detect suicidal behavior due to coding practices, reimbursement patterns, issues concerning ethics and safety, and the uncertainty of the patient's intent.
Nevertheless, most of the clinically relevant data in EHRs, such as signs and symptoms and condition severity, are frequently available in narrative text from MHPs but not in structured and coded form. 
For instance, most clinically relevant information on mental health conditions, such as depression and suicidal ideation, is available in unstructured clinical notes.
However, understanding and preventing suicide at an early stage requires data collected in real-time or through a source where individuals can express their life events and mood-related symptoms without the fear of social stigma. Social media platforms (e.g., Twitter, Reddit) can provide a rich source of insights on linguistic, interactional, and expressiveness features, complementing and supplementing clinical notes and interviews \cite{Gaur2018}. Similarly, the information derived from social media posts created by those experiencing suicidal ideation or attempt suicide may differ from what is typically documented in EHRs by MHPs, given the freedom of expression and its timeliness.  For instance, a user makes the following post on r/SuicideWatch: ``Really struggling with my \textit{bisexuality}, which is causing chaos in my \textit{relationship} with a girl. Being a fan of the LGBTQ community, I am equal to \textit{worthless} for her. I'm now starting to \textit{get drunk} because I can't cope with the \textit{obsessive}, \textit{intrusive thoughts}, the \textit{need to isolate myself}, and \textit{sleep forever}''. Note its clinical relevance: it signals \textit{Drug Abuse}, \textit{Obsessive Compulsive Disorder}, \textit{Suicidal Ideations}, and \textit{Borderline Personality Disorder} (BPD).  
While the information on suicidality derived from unstructured EHR text can support point-of-care clinical decision making for suicide prevention, social media text can provide additional perspectives for public health interventions.  
In our research, we seek to demonstrate the supplementary and complimentary relationships between EHR and social media in recognizing individuals at risk of suicide. Leveraging the list of suicide risk factors (SRFs) identified by Jashinsky et al. \cite{Jashinsky2014}, we showcase the similarities and dissimilarities in their manifestation on social media and EHR data. For this task, we utilize social media and EHRs from Reddit and Weill Cornell Medicine Ambulatory EHR clinical notes (WCM EHR), respectively. 
Recent research on identifying users with depression \cite{Guntuku2017}, estimating the severity of mental illness \cite{Gaur2018} or analyzing the change in user's expression as they change topics of their conversation depending on their current conditions, mental health communities  (or MH-subreddits) on Reddit have been effective in gleaning actionable insights. The creation of communities specific to mental illnesses on Reddit (e.g., r/Depression, r/Autism, r/PTSD, r/SuicideWatch) has facilitated clinical inferencing because of flexibility in the length of the post and the trust because of human moderation. Such characteristics have made Reddit provide nuanced content over Twitter. 
WCM EHRs are the unstructured and heterogeneous clinical notes are rich in content with granular details related to suicide risk. The user and content distributional similarity between Reddit and WCM EHR actuated our study to compare and contrast mentions of SRFs in both the platforms (see Datasets section). 

In this study, we extract and compare SRFs derived from unstructured clinical text data and Reddit. While we mainly focus on user postings in r/SuicideWatch, we identify and gather semantically similar postings made by the user in other MH-subreddits. In this study, we label these emerging SRFs as ``Other Important SRFs''. Considering these key insights, we discuss the complementary or supplementary relationship between r/SuicideWatch and WCM EHRs in discerning expressions of SRFs. Subsequently, we report the limitations of our analytical study as a recommendation to improve future research focusing on associating clinical settings and social media for prompt assessment of suicidality of an individual.  

Suicide risk prediction from unobtrusively gathered and up-to-date social media data has been beneficial in understanding suicide-related behaviors of users suffering from mental health conditions \cite{Alvarez-JimenezM2013ipoliticaf, Jashinsky2014}. A study by Alvarez et al. showed that mediated social media-based therapy could successfully support early interventions for patients with the first episode of psychosis \cite{Alvarez-JimenezM2013ipoliticaf}. Platforms such as Twitter, Reddit, and Facebook have provided data relevant to depression, suicidality, postpartum depression, and post-traumatic stress disorders \cite{Merchant2019}. Twitter has been investigated for depression symptoms, suicide ideations, and SRFs for potential insights on early intervention in emergency \cite{Luo2020}. Comparing the patient's perspectives gleaned from social media can suggest suicide risk factors that can supplement and complement the findings from EHR \cite{Velupillai2019,gaur2019shades}. Merchant et al. conducted a study on $\sim$1000 consenting patients on Facebook suffering from 21 categories of mental conditions, including anxiety, depression, and psychosis, to show the utility of Facebook language as the screening tool estimate the onset of disease and conduct early interventions. The study concluded that contrasting insights from social media with EHR is necessary to leverage the findings for clinical use \cite{Merchant2019}. 

On the other hand, Roy et al.\cite{Roy2020} and Gaur et al.\cite{Gaur2020a} underlined that prior studies had ignored the use of clinical guidelines in their research, curtailing adoption of methods to practice . Studies from Chen et al. \cite{ChengPhD2017}, Howard et al. \cite{Howard2019} statistically explored natural behavioral language processing methods (e.g., use of first-person pronouns, sentiments, emotions, language models) and general lexicons (e.g., Linguistic Inquiry and Word Count (LIWC)) to study suicide risk without entailing clinical knowledge in the form of either medical resources or subject matter expert. We improve upon the limitations of previous work by incorporating relevant clinical information in the way of lexicons and medical knowledge bases in mental health care \cite{Gaur2020}.

\section*{Datasets} \label{sec:D}
The posts explicitly and implicitly expressing SRFs were extracted from Reddit using a set of domain-specific keywords that describe each SRFs independently. For instance, Depressive Feelings was expressed with following set of keywords: \textit{thoughts, emotions, ranting, hopeless, ocd}, 
Self Harm was expressed with following set of keywords: \textit{cuts, hurt, pills, overdose, tear, knife}. A more detailed list of keywords in provided on this \href{https://drive.google.com/file/d/1EcTEtXfpUSVXWR8vM5U8eg61kyuyhotL/view?usp=sharing}{\underline{link}}. The content covers  2,624,846 unique users and 2,469,893 posts across 15 mental health subreddits (MH-subreddits) (Bipolar, Borderline Personality Disorder, Depression, Anxiety, Opiates, Opiates Recovery, Self Harm, Stop Self Harm, BipolarSOs, Addiction, Schizophrenia, Autism, Aspergers, Crippling Alcoholism, BipolarReddit, SuicideWatch) over 11 years (2006-2016). This study focuses on users who have expressed suicidal tendencies on r/SuicideWatch and other mental health-related subreddits. We consider postings of a r/SuicideWatch in other subreddits as transient because they implicitly reflect a user's mental health status. As a consequence, we obtain posts with words related to some SRFs but posts are not related to SRFs. For instance: ``People accidentally cutting while shaving'' will be mis-labeled with following SRF ``injury of unknown intent'', if words like ``hair'', ``shave'', ``slack'', ``accidentally'' were not used to filter out. To filter such posts, we use a list of exclusion terms to prevent false positives while labeling posts with SRFs.  
A complete list of exclusion terms is provided on this \href{https://drive.google.com/file/d/116HHnrXhilwcM6hB7PYvreKQ2ttopJQ5/view?usp=sharing}{\underline{link}}. Considering this as a proxy of understanding co-morbidity on Reddit, it is essential to identify a user's content in other subreddits and measure similarity with content often posted by other users in r/SuicideWatch. We followed a quantitative procedure, termed as \textit{semantic relatedness} (a variant of cosine similarity measure). Thus, our final dataset contains 416,154 posts from 195,836 users with an average of $\sim$460 words per post, that is significantly larger and substantial than twitter dataset in Jashinsky et al. \cite{Jashinsky2014}. The Reddit dataset is available for download on this \href{https://drive.google.com/drive/folders/1nkxOM2FbHZr2fk181UM_meUMoI0nLX-G?usp=sharing}{\underline{link}}. 

For the EHRs data, we used the EpicCare® Ambulatory EHR platform (used by Weill Cornell Medicine's (WCM)). The platform documents clinical care in its outpatient settings, which constitutes the EHR data used in this study. In particular, we extracted all clinical notes for n=123,703 patients who either had a diagnosis of major depressive disorder or have been prescribed an antidepressant between 2007 to 2017.  Our corpus of all clinical records comprising more than 13.8 Million documents was authored by clinicians from multiple specialties, such as internal medicine, psychiatry, anesthesiology, pain medicine, across WCM outpatient clinics.  Notes were heterogeneous in their content and level of detail and unstructured in their format.  We subsequently generated representation of these datasets using word embedding models, which have shown to capture each word's meaning in context and derive clusters of semantically related words and phrases \cite{Mikolov2013 , Kingrani2017}. Finally, a comparative analysis was performed to meaningfully probe auxillary relationship between clinical and social media setting to understand the suicide risk factors. 

\begin{figure}[!htbp]  
  \begin{center}
    \includegraphics[width=75mm, scale=1.0, trim=1.0cm 2.5cm 2.0cm 2.5cm, angle=-90]{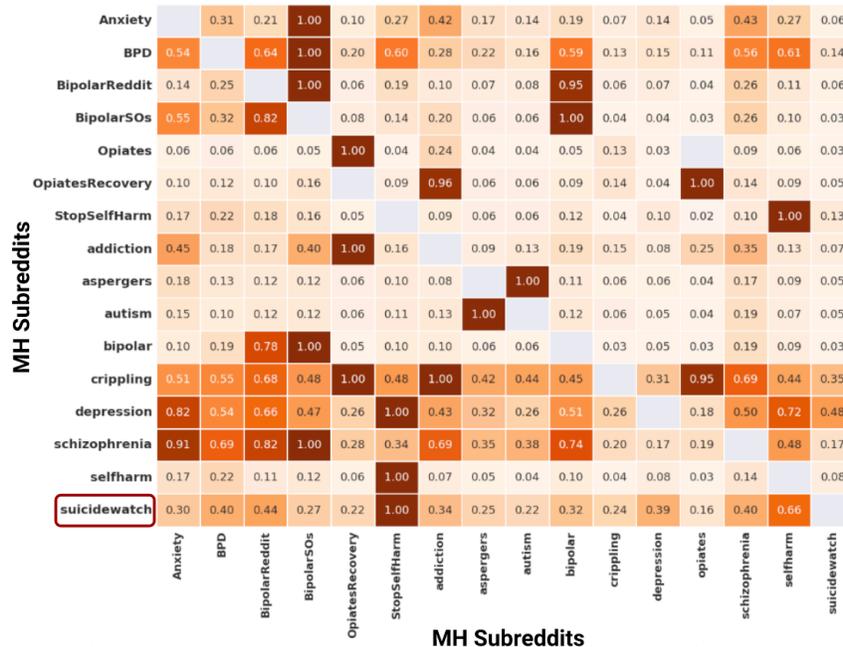}
  \end{center}
  \caption{Heatmap showing semantic relatedness between mental health subreddits based on user overlap. For instance, r/SuicideWatch and BPD has a score 0.40, which signifies,
  of the number of users in common to both subreddits, 
  content of 40\% of users overlaps. Semantic relatedness is measured following equation \ref{eq:1}. We have ignored comments in subreddits as they added minimal information gain.}
  \label{fig:subredditmap}
\end{figure}

\section*{Methods} \label{sec:M}
We develop a method to quantify the semantic relatedness of the content produced by a user in r/SuicideWatch and other mental health subreddits (see Datasets section). A fundamental challenge that social media platform raise is false positives.  A segment of resilient users on r/SuicideWatch shares their experiences to support others. Recognizing and separating supportive users from potentially suicidal users is essential for a reliable comparison with clinical notes. We utilize the suicide risk severity lexicon built using the Columbia-Suicide Severity Rating Scale (C-SSRS) to eliminate supportive content \cite{Gaur2019}. Consequently, we build an SRF-related lexicon as a composition of relevant lexicons created in past studies to recognize medical concepts in r/SuicideWatch subreddit and clinical notes \cite{Gaur2019}. Such a process is termed as entity (or concepts in lexicon) normalization and it requires a numerical representation in the form of a vector of length $\mathcal{L}$ ($V_{\mathcal{L}}$) (where $|V_{\mathcal{L}}|$ can be of either dimensions \{300, 200, 100, 50\} and $V_{\mathcal{L}} \in \mathbb{R}$). This numerical vector representation or embedding of the word are generated using a word embedding model \cite{Gaur2019}. Since our task is specific to mental health and suicide risk, we require fine-tuning the word embedding model. For this purpose, we utilize SRF-related lexicon. After that, we used a non-parametric clustering approach to cluster embeddings and associate SRFs with them based on their semantic similarity (details in the Methodology section). On the analyzing the clusters labeled with one or set of SRFs, we found that SRFs such as ``gun ownership'' and ``suicide around individual'' are often mentioned together in clinical notes, suggesting that gun was the medium for suicide.
On the other hand, ``gun ownership'' frequently occurred with SRFs such as ``depressive feelings'', ``psychological disorders'', ``suicide ideations'' in the r/SuicideWatch. It suggests that SRFs: ``depressive feelings'', ``psychological disorders'', ``suicide ideations'' cause owning the gun. Hence, complementing the findings of clinical notes with Reddit would provide a better picture of the severity of an individual's suicide risk. Our analysis showed that the list of SRFs stated in Jashinsky et al. is not sufficient for a complete comparison of the two platforms. For instance, SRFs such as ``poor performance in school'', ``relationship issues'', ``racial discrimination'' are major contributors of suicide ideations but are not specified in Jashinsky et al.'s list of 12 SRFs: ``depressive feelings'', ``depression symptoms'', ``drug abuse'', ``prior suicide attempts'', ``suicide around individual'', ``suicide ideation'', ``self-harm'', ``bullying behavior'', ``gun ownership'', ``psychological disorder'', ``family violence and discord'', and ``impulsivity''  \cite{Jashinsky2014}. 
It is measured between two sub-reddits as the overlap in content made by users in common to both the subreddit. In our study, we formalize semantic relatedness (SR($S_1$, $S_2$)) between two subreddits ( $S_1$ and $S_2$) as follows:

\begin{equation} 
\label{eq:1}
\mbox{SR}(S_1, S_2) = \frac{\sum_{u \in S_1, S_2}\frac{\sum_{p_i \in \mbox{posts}(S_1), p_j \in \mbox{posts}(S_2)} \mathbf{\delta}(\vec{u}_{p_i}^{S_1},\vec{u}_{p_j}^{S_2})} {|\mbox{posts}(S_1)|+|\mbox{posts}(S_2)|}}{\mbox{N}_u}; 
\mathbf{\delta}(\vec{v}_x, \vec{v}_y) = \begin{cases}
1 ; \mbox{cos}(\vec{v}_x, \vec{v}_y) > 0.9 \\
0 ; \mbox{otherwise}
\end{cases}
\end{equation}
where $\vec{u}_{p_i}^{S_1}$ is the vector representation of a post ($p_i$) made by a user ($u$) in a subreddit ($S_1$) and $N_u$ is the number of users common to both the subreddits. The threshold for the similarity is 0.9, which is empirically defined based on domain expert judgment. We follow this process over all the MH subreddits, as shown in Figure \ref{fig:subredditmap}.  From Figure \ref{fig:subredditmap}, an SR score of 1.0 between r/SuicideWatch and r/StopSelfHarm suggest that users in common to these subreddits have more similar content compared to  r/SuicideWatch and r/Opiates (SR score = 0.16). Based on the threshold of 0.40 set on the SR score, we extracted and gathered posts of r/SuicideWatch users in Depression, Addiction, Anxiety, Bipolar, Stop Self Harm, Self Harm, Borderline Personality Disorder (BPD), and  Schizophrenia subreddits. The aggregated suicide-related content contains posts which have negations and conjunctions. Generating embeddings of these posts is erroneous as its is difficult to generate a semantic-preserving representations of posts with negations and conjuctions. Thus, we identified these posts and remove them for the study. Further, we leveraged a suicide risk severity lexicon (See details under Methods) to filter our posts which are not suicide-risk-related. With this pre-processing method we identified a cohort of  2180 (2.5\% of 93K, $\sim$100K posts) users who were potentially suicidal through expressions of suicide risk factors and associated mental health conditions \cite{Alambo2020}. We extracted the content of these users in other MH subreddits and aggregated to create the dataset for the study. The reliability of the dataset was evaluated through an annotation performed over a randomly sampled 500 users ($\sim$30K Posts). The annotation was performed by psychiatrists using 5-labels: Supportive, Indicator, Ideation, Behavior, and Attempt, of which \{Ideation, Behavior, Attempt\} are defined in the Columbia-Suicide Severity Rating Scale.  The content was annotated at the user-level and at the post-level. The inter-rater reliability score was recorded through pairwise and groupwise agreement using the Krippendorff metric \cite{Yaseen2019}. 
Pairwise agreement is conducted between pairs of annotators and the annotator with high agreement score is selected for groupwise agreement. In this annotation agreement scheme, the annotations of the selected annotator is compared with mutually agreed annotations from an incremental group of annotators ( in our case \{2,3\}). If there is a substantial agreement between the selected annotator and other groups of annotators with varied sizes, we consider the annotation, else process is repeated with next best annotator in pairwise scheme. Both the agreement schemes, together achieve robustness in the annotation task. Both at the user-level and post-level, we obtained a substantial inter-rater reliability score by  measuring pairwise and groupwise agreement. At user-level, pairwise agreement was 0.79 and groupwise agreement was 0.69. In post-level, pairwise agreement was 0.88 and groupwise agreement was 0.76.

\begin{figure*}[!htbp]  
  \begin{center}
    \includegraphics[width=55mm, scale=1.0, trim=6.0cm 4.0cm 6.0cm 3.5cm, angle=-90]{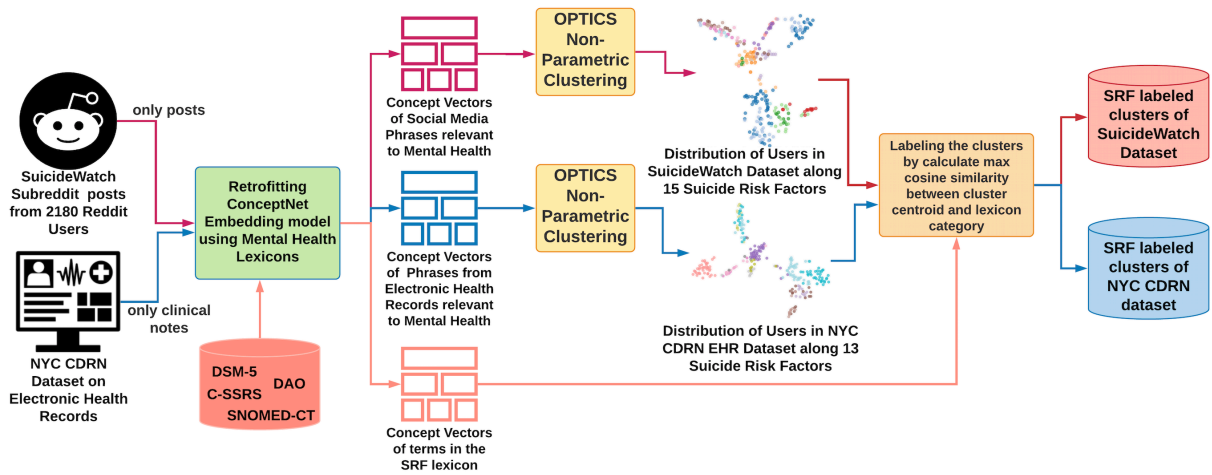}
  \end{center}
  \caption{Overall workflow of an unsupervised approach to understand suicide risk factors (SRFs) in r/SuicideWatch posts and WCM EHR clinical notes using ConceptNet and semantic lexicons.}
  \label{fig:architecture}
\end{figure*}

We describe an unsupervised and clinically grounded SRF-labeling methodology to identify and compare the different SRFs expressed in the voluminous r/SuicideWatch posts and clinical notes in EHRs. The proposed methods inputs sentence-level embeddings of the posts and clinical notes, and word embedding of the concepts in the SRF lexicon. The outcome, independent clusters of r/SuicideWatch posts, and clinical notes were associated with an SRF or set of SRFs by measuring the similarity between the embeddings of the centroid of the clusters and SRFs. The common and disparate SRFs were identified from the two platforms and compared (see Figure \ref{fig:architecture}).

To identify SRF-related concepts (words or phrases) from online conversations on r/SuicideWatch, we require semantic lexicons created specifically for suicide risk. We employ lexical-resources based on Columbia Suicide Severity Rating Scale (C-SSRS) \cite{Yaseen2019} and Diagnostic and Statistical Manual of Mental Disorders (DSM-5) developed by Gaur et al. \cite{Gaur2019}. We used the concepts in the created SRF lexicon as the seed concepts to enrich it with terms in SNOMED-CT, and Drug Abuse Ontology (DAO) \cite{Gaur2019} following a guided markovian random walk procedure.
An instance of the created SRF lexicon associates `` Suicide Ideations'' with ``intrusive thoughts''. To verify this association, we trace the path from ``Obsessive-compulsive disorder'' ([SNOMEDCT: 1376001]) to ``intrusive thoughts.'' Obsessive-compulsive disorder [SNOMEDCT: 1376001] is associated with Suicidal Ideations [SNOMEDCT: 425104003], and the concept ``Intrusive Thoughts'' [SNOMEDCT: 225445003] is a child of parent concept ``Disturbance in Thinking'' [SNOMEDCT: 26628009], which is a child concept of ``Obsessive-Compulsive Disorder.'' Thus proving the association between ``intrusive thoughts''  and ``Suicidal Ideations''. The SRF  lexicon acts as a component to fine-tune a generic word embedding model, ConceptNet \cite{Yaseen2019}, to generate contextualized representations of r/SuicideWatch posts and clinical notes in WCM EHRs. This post-processing technique is called retrofitting, and it reinforces the embedding of words by minimizing the distance between concepts that are relevant in describing SRFs \cite{Alambo2020}. For example, in the retrofitted ConceptNet, ``hopeless'' and ``depressive feelings'' are in proximity compared to ``hopeless'' and ``harassment.'' The proximity suggests that depressive feelings are expressed with the term "hopeless" more often than "harassment" in suicide risk-related conversations.  Another example is the semantic proximity of ``impulsivity'' to ``bullying'' rather ``family violence and discord.'' 

After retrofitting of ConceptNet embedding model, we leverage it to generate vector representations of each post in the suicide dataset created from r/SuicideWatch and other relevant online mental health communities.  Similarly, we create representations of clinical notes documented in EHRs. Note that our method to create representation is post-level and clinical notes-level, not user-level. Psychiatrists treat a siloed community of patients suffering from mental health disorders, which restricts diversification. A strategic comparison of clinical notes in EHRs with population-level social media markers could enable the psychiatrists to develop better contextual questions in diagnostic interviews and elucidate disease epidemiology for better patient engagement. The representations of the two sources of content were clustered independently using a non-parametric clustering algorithm, OPTICS (Ordering Points To Identify the Clustering Structure) \cite{Luo2020}. Our selection of OPTICS over approaches such as DBScan, K-Means, Gaussian Mixture Models, is based on the clustering algorithm's ability to create diverse (at least equal to the number of SRFs) clusters, where each cluster most-likely cohesively represents an SRF. We calculate the similarity between the representation of the centroid of the cluster and the SRFs. The SRF, with the highest similarity with the centroid, is the estimated label of the cluster. We followed this process to label clusters created from suicide dataset and clinical notes in EHRs.

\section*{Results and Discussion} \label{sec:Dis}
\begin{figure}[ht]
\centering
\begin{subfigure}{.5\textwidth}
  \centering
  \includegraphics[width=35mm, scale=1.0, trim=5.cm 6.5cm 4.0cm 1.0cm, angle=-90]{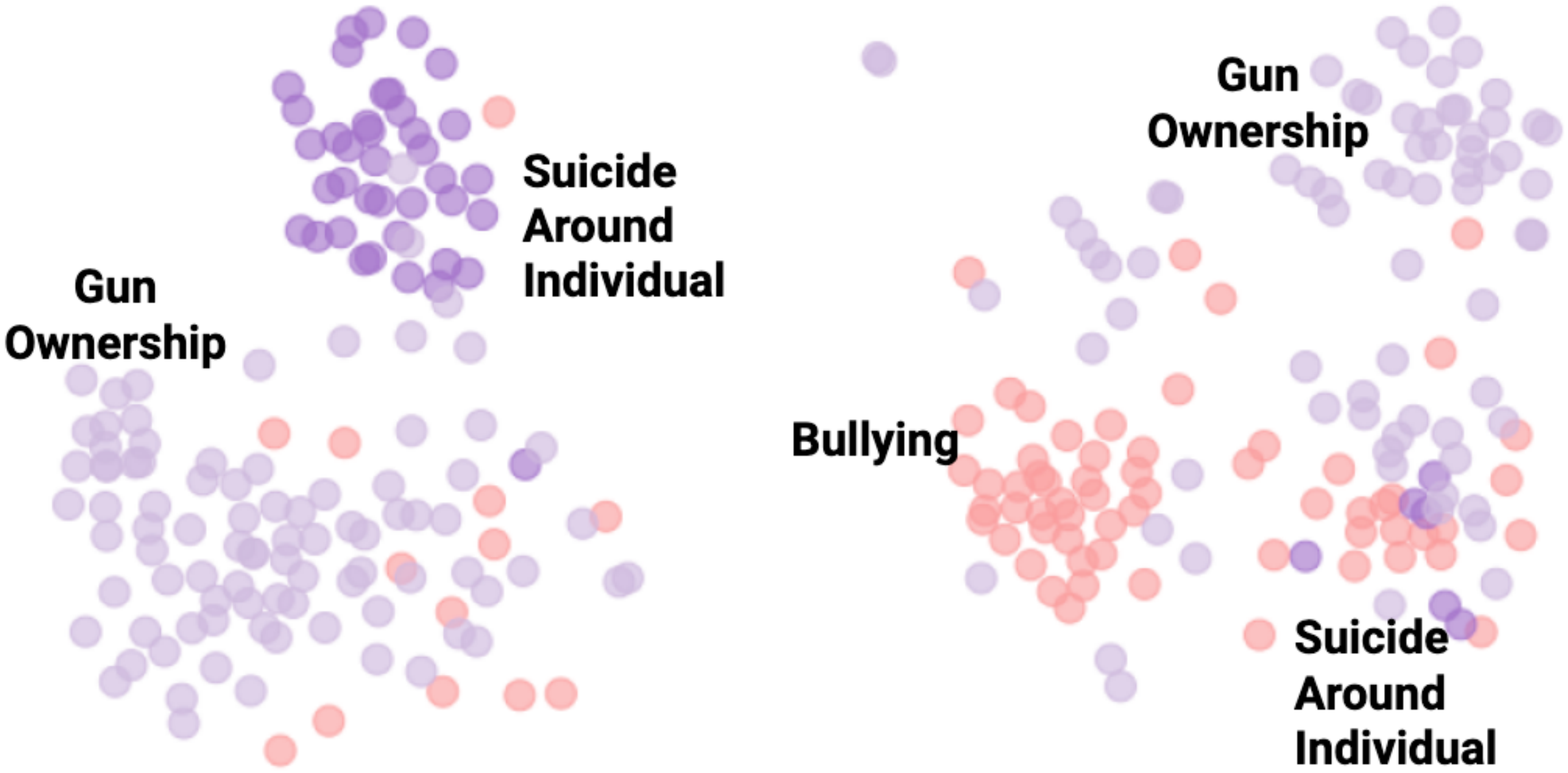}
  \caption{Top three SRFs elicited from clustering clinical notes in WCM EHR dataset. SRF such as Suicide around individual is often mentioned with gun ownership and bullying behavior.}
  \label{fig:ehr_top_trip}
\end{subfigure}%
\hspace{0.1cm}
\begin{subfigure}{.5\textwidth}
  \centering
  \includegraphics[width=55mm, scale=1.0, trim=3.5cm 6.5cm 6.0cm 1.0cm, angle=-90]{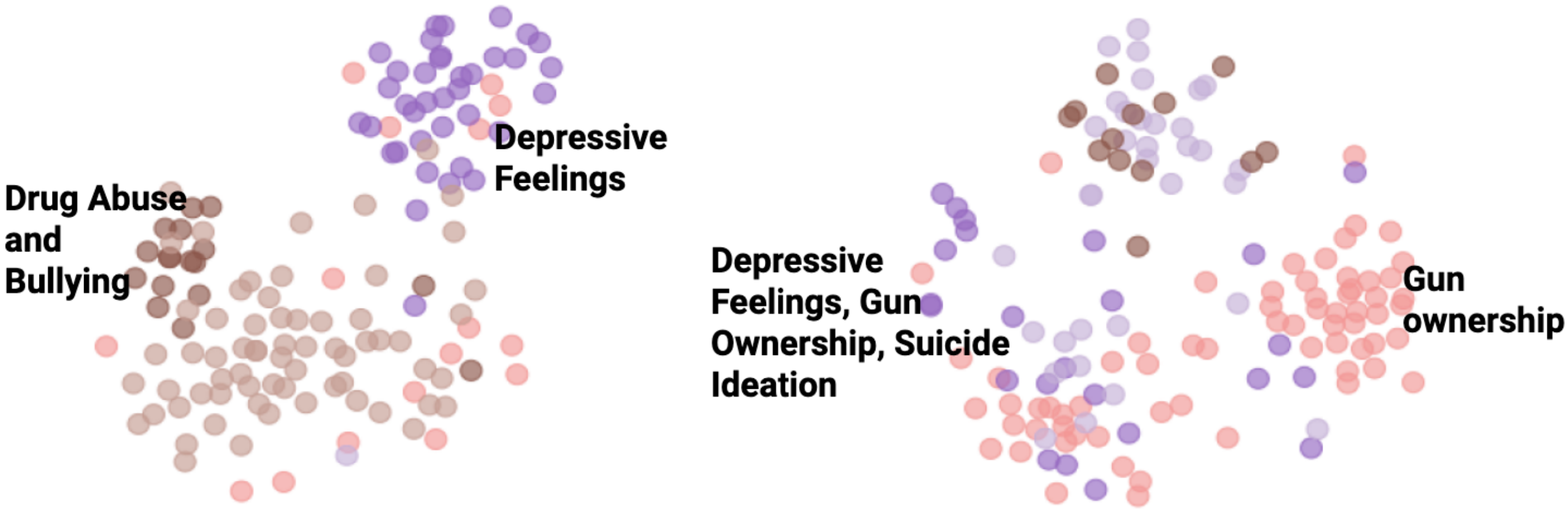}
  \caption{Top five SRFs elicited from clustering r/SuicideWatch posts. Depressive feelings, bullying behavior, and drug abuse are often mentioned together. Suicide ideations are expressed by users mentioning signs of depressive feelings and threatening to possess gun.}
  \label{fig:reddit_penta}
\end{subfigure}
\caption{Description of top clusters showing interrelation between different SRFs identified in r/SuicideWatch  and clinical notes.}
\label{fig:top_cluster_EHR_reddit}
\end{figure}

\begin{figure}[!ht]
\centering
\begin{subfigure}{.5\textwidth}
  \includegraphics[width=70mm, scale=1.2, trim=1.0cm 2.0cm 0.5cm 1.0cm,
  angle=-90]{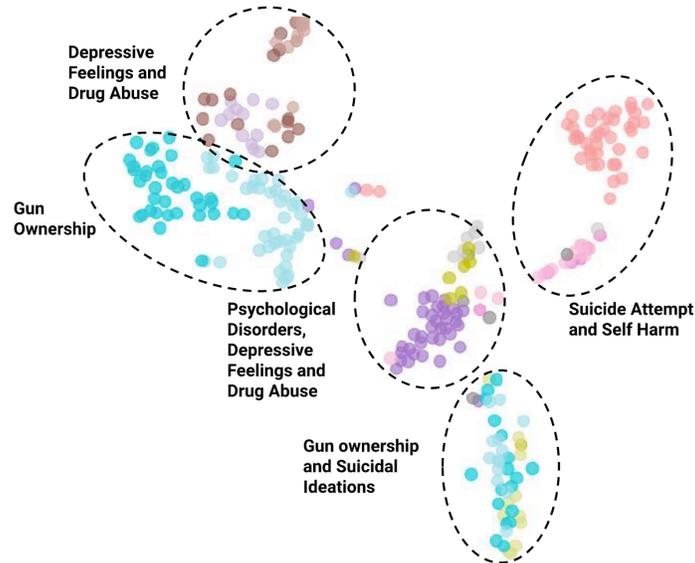}
  \caption{SRF clusters of clinical notes. Out of 14, 7 SRFs were identified from clinical notes.}
  \label{fig:ehrv2}
\end{subfigure}
\begin{subfigure}{.5\textwidth}
  \includegraphics[width=75mm, scale=1.2, trim=0.5cm 1.5cm 1.0cm 1.0cm,
  angle=-90]{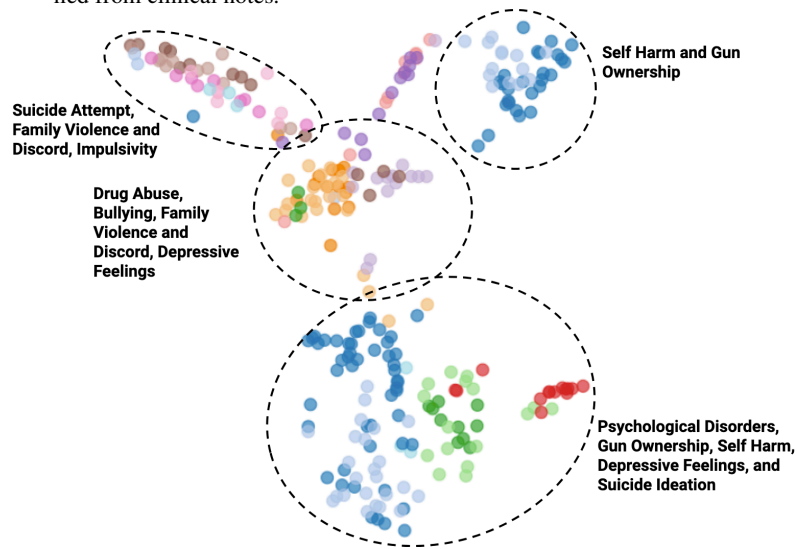}
  \caption{SRF clusters of r/SuicideWatch  posts. Out of 14, 10 SRFs were identified from the posts made by users on r/SuicideWatch.}
  \label{fig:redditv1}
\end{subfigure}
\caption{Clusters of SRFs identified from clinical notes and r/SuicideWatch  posts.}
\label{fig:cluster_EHR_reddit}
\end{figure}

On the clusters labeled with a set of SRFs, this study discusses the commonalities and differences in the expressions of suicide-risk from patients and users in WCM EHRs and r/SuicideWatch respectively. For instance, \textit{depressive feelings, psychological disorders, drug abuse, and suicide ideations} are the common SRFs communicated on both platforms. However, both the platform differs from each other concerning following SRFs: \textit{depressive symptoms} and \textit{suicide around individual} is revealed only from clinical notes; \textit{bullying behavior, self-harm, impulsivity}, and \textit{family violence and discord} significantly manifests in r/SuicideWatch communications only. We ranked the SRFs independently for each platform. In clinical notes, most frequent SRFs are depressive feeling (24\%), psychological disorders (21.1\%), drug abuse (18.2\%), depressive symptoms (14.9\%), suicide around individual (12.6\%), and suicide ideations (9.1\%) (see Figure \ref{fig:ehrv2}). On analyzing the clusters derived from the EHR data, we observed mentions of gun ownership contextualizes bullying behavior and suicide around individual (see Figure \ref{fig:ehr_top_trip}).

In r/SuicideWatch posts, gun ownership (17.4\%), self-harm (14.6\%), bullying behavior (13.2\%), drug abuse (13\%), depressive feelings (11.6\%), suicide ideation (10.7\%), psychological disorders (10\%), impulsivity (9.6\%) are frequently discussed (see Figure \ref{fig:reddit_penta} and Figure \ref{fig:redditv1}). A semantic analysis of the posts on r/SuicideWatch showed family violence and discord as the reason for impulsivity, leading to a suicide attempt (see Figure \ref{fig:redditv1}). Further, suicidality measured through co-occurrence of drug abuse and bullying behavior often showed a high frequency of terms mapped to family violence, then depressive feelings.

However, we seldom saw the manifestation of SRF, ``family violence and discord'' in the documented clinical notes. Clinical settings steered away from discussing family violence and impulsivity, significant SRFs mentioned on \\ r/SuicideWatch. This may indicate that individuals who act impulsively, leading to the risk of committing suicide, are less inclined to discuss such behavior privately with their MHPs. On the contrary, Depressive feelings were prominent SRF followed by psychological disorders and drug abuse in clinical settings. Figure \ref{fig:ehrv2} shows that semantic clusters of suicidal ideations and gun ownership were formed from the representations of clinical notes, which is consistent with the clusters derived from r/SuicideWatch and is illustrated in Figure \ref{fig:redditv1}. Further, written communications from users on r/SuicideWatch mention suicide ideations when describing psychological disorders and guns as threats to self-harm. These findings are relatable with clinical insights derived from C-SSRS \cite{Yaseen2019}. 

Both the platforms show content from users (or patients) discussing gun ownership and bullying behaviors when expressing potential suicide risk. In contrast, r/SuicideWatch clusters were more revealing by co-locating other SRFs such as drug abuse and self-harm. Considering the findings from the analysis of the suicide risk-related content this study specify a supplementary and complementary relationship between r/SuicideWatch and EHRs. Besides, while our clinically grounded approach demonstrated the feasibility to intervene among those at suicide risk consensually associating their social media profile with EHR data, we plan to elucidate certain computational and practical limitations in our future research.

\section*{Limitations} \label{sec:L}
The findings described in the study should be interpreted in the context of some limitations concerning (a) the semantic lexicon, (b) the suicide dataset, and (c) the WCM EHR dataset.
The abstraction of posts made on r/SuicideWatch using the semantic lexicon comes with a limitation regarding its completeness. The lexicon utilized in this study is a composition of concepts in PHQ-9, C-SSRS, DAO, and SNOMED-CT, which was semantically appropriate for this study. But it can be improved with slang terms and moderation from domain experts. However, care must be taken while extending the lexicon as it brings ambiguity which might falsely determine SRFs. The suicide dataset prepared by strategically accumulating the content from MH-subreddits ignored some subreddits directly related to r/SuicideWatch but are sparse. For example, r/euthanasia (assisted suicide) and r/suicideprevention, are other discussion forums on suicide risk that were not included in this study. 

In addition to the list of SRFs provided by Jashinsky et al. \cite{Jashinsky2014}, we found additional topics such as ``relationship issues'', ``brain damage'', ``physiological stressors'', ``cant afford rent, debt, failure'', and ``unemployment'' that are significant contributors to suicide ideations but could not be mapped to the existing list and require a clinically relevant SRF label. For now, we considered these stressors as "Other Important SRFs," and in this study, we did not provide a comparison between the two platforms based on this category. This is because it is a mixed category in terms of SRFs; hence a vector representation would be semantically misleading, causing false inferences. Likewise, we formed another category, termed as ``Accessory,'' which contains phrases having mention of a material or substance that assisted suicide. For instance ``self-inflicted injury by suffocation by plastic bag'' (a plastic bag is an accessory), ``suicide or self injury by jumping from bridge'' (the bridge is a navigational concept [SNOMEDCT: 242843002]), ``suicide or self injury by caustic substance'', ``attempt suicide by car exhaust (event is an accessory)'', ``indirect self harm due to mechanical threat'' (trapped in a car trunk, refrigerator, etc.). Like ``Other Important SRFs'', ``Accessory'' is a mixed category, we could not assess the commonalities and disparities between r/SuicideWatch and WCM EHR clinical notes. As future work, we will explore these incohesive categories from a clinical perspective and further strengthen our study with demographic and spatial information. Additionally, we will explore ways to differentiate between suicide completers and suicide attempters.

\section*{Conclusion}
Suicide risk factors can be determined and used for suicide prevention at an early stage; however, poorly documented clinical notes curtail MHPs from devising intervention strategies. Further, EHRs shed some light on a patient's current and anew psychopathology status, but fall to cultivate a broader understanding of the mental health conditions. Recently, people have investigated social media, mainly Reddit, to gather insightful population-level markers for assisting MHPs. However, ambiguous and sparse content in Reddit and EHRs requires structured hierarchical knowledge for apprehension and effective decision making. In this work, we investigated the commonality and disparity in the conversations specific to SRFs from users on r/SuicideWatch and patients in the clinical setting. In the process, we created an SRF-specific lexicon for semi-automatically 
identifying medical concepts on r/SuicideWatch and clinical notes for contextualization and semantic clustering of SRFs.  We observed a few similarities between the SRFs discussed within the private EHR data versus an anonymized, public setting, SW. Simultaneously, many dissimilarities were observed across the datasets suggesting future studies should focus on linking clinical and non-clinical data at an individual level to get a comprehensive view of an individual's suicide risk. 
The post-level and user-level annotated dataset created from r/SuicideWatch will be made publicly available upon acceptance of the study. Further, the source code developed to conduct this study will be made online on github for reproducibility.  

\section*{Acknowledgement}
This work was funded in part by NIH grants R01MH105384, R01MH119177, and P50MH113838.  Any opinions,  findings,  and conclusions/recommendations expressed in this material are those of the author(s)  and do not necessarily reflect the views of the NIH.

\bibliographystyle{unsrt}
\bibliography{reference}

\end{document}